# EFFECT OF Mo-SUBSTITUTION ON SUPERCONDUCTIVITY, FLUX PINNING AND CRITICAL CURRENTS OF $La_{1.5}Nd_{0.5}Ca_1Ba_2Cu_5O_z$

S. Rayaprol[1+], Krushna Mavani[1], C. M. Thaker[1], D. S. Rana[1], R. G. Kulkarni[1]
D. G. Kuberkar[1*], Darshan Kundaliya[1+] and S. K. Malik[+]

[1] Department of Physics, Saurashtra University, Rajkot – 360 005 (INDIA)
[+] Tata Institute of Fundamental Research, Mumbai – 400 005 (INDIA)

[*] Corresponding Author

Dr. D. G. Kuberkar
Associate Professor

| | |
|---|---|
| Address for Correspondence | Department of Physics<br>Saurashtra University<br>Rajkot – 360 005<br>INDIA |
| Phone Number | +91-281-571811 |
| Fax Number | |
| Email | dgk_ad1@sancharnet.in / dgk@icenet.net |

**ABSTRACT**

Mo-substituted triple perovskite superconducting samples with stoichiometric composition $La_{1.5}Nd_{0.5}Ca_{1}Ba_{2}(Cu_{1-x}Mo_{x})_{5}O_{z}$ with x = 0.00, 0.01, 0.03, 0.05 and 0.10, have been investigated using X-ray diffraction, electrical resistivity, magnetic susceptibility and Magnetization measurements. Oxygen content has also been determined by the iodometric technique. The observed reduction in superconducting transition temperature, $T_c$, in this system with increasing Mo content can be attributed to the decrease in hole concentration. Magnetization measurements as a function of temperature and field (H), exhibit an M-H hysteresis loop which indicates moderate flux pinning, which is thought due to pinning centers created by Mo substitution at Cu sites. The critical current density, $J_c$, deduced from the M-H data satisfies the flux creep model.

## 1. INTRODUCTION

The superconducting system $La_{1.5}R_{0.5}Ca_1Ba_2Cu_5O_z$ with R = rare earth (abbreviated as La-2125) has been extensively studied by our group [1, 2]. La-2125 has a triple perovskite structure similar to that of the tetragonal $La_{2.5}R_{0.5}Ca_1Ba_3Cu_7O_z$ (La-3137) [3] and a superconducting transition temperature ($T_c$) of ~78 K. Both La-2125 and La-3137 systems originate from the addition of rock salt like "$CaCuO_2$" layers in non-superconducting $La_2Ba_2Cu_4O_z$ and $La_3Ba_3Cu_6O_z$ compounds respectively.

The effect of substitution by various elements on the physical properties of $RBa_2Cu_3O_z$ (R=Y or rare earth ion, R-123) superconductors has been studied to investigate the origin of superconductivity in such compounds. The valency of the substituted element plays an important role in determining the effective copper valence (the mobile carrier concentration, p) or the oxygen content of R-123, both of which being crucial for the occurrence of superconductivity in such compounds. The substitution of $Fe^{3+}$, $Co^{3+}$, $Ni^{2+}$, $Zn^{2+}$, $Hf^{4+}$, $Mo^{4+,6+}$ etc. at the Cu site in R-123 reduces the $T_c$ [4, 5] while rare-earth substitution at the nominal Y site shows almost no effect on superconductivity [6]. Partial rare-earth substitution for La in $La_{2-x}R_xCa_yBa_2Cu_{4+y}O_z$ also exhibits almost no effect on superconductivity [7], similar to the situation in rare-earth substituted R-123 [6]. To date, to the best of our knowledge, no attempts have been made to substitute elements such as Fe, Co, Ni, Hf, Mo etc. for Cu in $La_{1.5}R_{0.5}Ca_1Ba_2Cu_5O_z$ (LRCBCuMO) to investigate the effects of such substituents on structure, superconductivity, critical current and flux pinning. We have initiated a systematic study of such substitutions and report here the results on Mo-substituted samples.

Since the valence state of Mo (4+ or 6+) is higher than that of Cu (2+), a substitution of Mo for Cu in LNCBCuMO (for R = Nd in LRCBCuMO), may decrease the carrier concentration p and hence $T_c$ due to either hole filling or pair breaking or both. The substitution of Mo at Cu site is also equivalent to the introduction of defects at Cu-sites, which may act as pinning centers. Therefore, it is of interest to investigate the effect of Mo-substitution for Cu in LNCBCuMO on oxygen content or p, superconductivity ($T_c$), flux pinning and $J_c$.

In earlier studies, some of us have investigated the effects of Mo substitution for Cu in $YBa_2Cu_3O_z$ [8, 9] and $GdBa_2Cu_3O_z$ [5] and have found structural transformation taking place from orthorhombic to tetragonal structure accompanied by decrease in $T_c$ and p with increasing Mo content. The aim of this paper is to study the effect of Mo doping for Cu on structural, superconducting, flux pinning and other physical properties of LNCBCuMO and compare the results with those reported for Mo doped R-123 systems [5, 8, 9].

## 2. EXPERIMENTAL

All the samples of $La_{1.5}Nd_{0.5}Ca_1Ba_2(Cu_{1-x}Mo_x)_5O_z$ series ($0 \leq x \leq 0.1$), were prepared by solid state reaction method. The high purity (>99.9%) starting compounds of $La_2O_3$, $Nd_2O_3$, $CaCO_3$, $BaCO_3$, CuO and $MoO_3$ were taken in stoichiometric quantities and thoroughly ground under acetone using agate and mortar. The mixtures were calcined at $930^0$C for 50 hrs with intermediate grindings. The resultant black powder was then palletized and sintered in the temperature range $940^0C$ – $950^0C$ for 24 hrs. These sintered samples were then annealed in flowing nitrogen gas for 12 hrs at $900^0C$, in flowing oxygen gas for another 12 hrs at the same temperature, then slow cooled ($1^0C$ / min) to $500^0C$ and kept in oxygen atmosphere for 24 hrs followed by slow cooling ($1^0C$ / min) to room temperature.

The X-ray diffraction patterns were recorded at room temperature using Cu-$K_\alpha$ radiation ($\lambda$ = 1.5405 Å). The superconducting transition temperatures were determined from the resistance measured by four-probe method. The magnetic susceptibility/magnetization, measurements were carried out on a Superconducting Quantum Interference Device (SQUID) magnetometer (Quantum Design) at TIFR, Mumbai. The oxygen content was determined by iodometric double titration method. From which hole concentration has been calculated.

## 3. RESULTS

Although it is rather difficult to predict the valence of Mo to be 4+ or 6+ on the basis of the present work, previous Mo-doped $YBa_2Cu_3O_z$ [8, 9] and $GdBa_2Cu_3O_z$ [5] studies have suggested a preference for 4+ valence for Mo, which we are assuming here.

The XRD patterns of $La_{1.5}Nd_{0.5}Ca_1Ba_2(Cu_{1-x}Mo_x)_5O_z$ samples are similar to the XRD patterns of La-2125 compounds [1, 2]. The lattice parameters, obtained from a least-square refinement of the d values observed from the X-ray diffractogram, are shown in Figure 1 as a function of Mo concentration. All the $La_{1.5}Nd_{0.5}Ca_1Ba_2(Cu_{1-x}Mo_x)_5O_z$ samples retain a tetragonal triple perovskite structure without transforming to the orthorhombic structure, whereas structural transition from orthorhombic to tetragonal structure is observed for the Mo doped Y-123 [8, 9] and the Y-123 superconductor [5] during the cooling process. The unit cell volume of $La_{1.5}Nd_{0.5}Ca_1Ba_2(Cu_{1-x}Mo_x)_5O_z$ compounds initially increases slightly for x = 0.0 to 0.03 and thereafter decreases with increasing x for x = 0.05 – 0.10 (Figure 2). The oxygen stoichiometry, z and z' and the corresponding hole concentrations, p for $La_{1.5}Nd_{0.5}Ca_1Ba_2(Cu_{1-x}Mo_x)_5O_z$ compounds are listed in Table – 1. Here z denotes the amount of oxygen per chemical formula of $La_{1.5}Nd_{0.5}Ca_1Ba_2(Cu_{1-x}Mo_x)_5O_z$ and z' is the amount of oxygen content per unit formula when nominally written in R-123 form as $(La_{0.3}Nd_{0.3}Ca_{0.4})(Ba_{1.2}La_{0.6}Ca_{0.2})(Cu_{1-x}Mo_x)_3O_{z'}$ (in the R-123 format with z' = 3/5*z). It has been observed for the La-2125 compounds that the hole concentration in sheets, $p_{sh}$, responsible for the superconductivity [1, 2]. Hence, the values of the hole concentration, p, and concentration of holes in Cu-O sheet, $p_{sh}$, were calculated by the Tokura model [11] using the oxygen content values obtained from iodometric analysis. It is evident from Figure 2 that the z and $p_{sh}$ decrease with increasing Mo content from x = 0.00 to 0.10 for $La_{1.5}Nd_{0.5}Ca_1Ba_2(Cu_{1-x}Mo_x)_5O_z$ series suggesting that $Mo^{4+}$ replaces $Cu^{2+}$ . Similar decrease in p and z with increasing Mo-content has been observed in Mo-doped Y-123 [8] and Gd-123 [5].

The electrical resistivity versus temperature plots for the $La_{1.5}Nd_{0.5}Ca_1Ba_2(Cu_{1-x}Mo_x)_5O_z$ samples are displayed in Figure 3 (a). Table – 2 lists the resistive superconducting transition temperatures, onset, mid-point and zero resistance. The $T_c^{R=0}$ is found to decrease from 79 K to 36 K on 10 % Mo substitution for Cu (Table – 2) with an average rate of ~ 4.3 K per At. % of Mo, seen clearly in Figure 4. The magnetic susceptibility of $La_{1.5}Nd_{0.5}Ca_1Ba_2(Cu_{1-x}Mo_x)_5O_z$ compounds is shown as a function of temperature in Figure 3 (b), and the magnetic onset transition temperatures $T_c^{ON}(\chi_{mag})$ listed in Table – 2 agree very well with the resistively determined $T_c^{R=0}$. Figure 5 shows typical magnetization (M) versus applied field (H) hysteresis loop indicative of flux pinning. It is customary to use magnetization measurements as a contact less method for determining the intragrain critical current density ($J_c$). From the width of the hysteresis loop, it is possible to estimate the $J_c$ in the superconducting portions of the samples using Bean's critical state model [12]. For a spherical grain of diameter D (in cm), $J_c$ (in A $cm^{-2}$) can be expressed in terms of magnetization M (in emu $cm^{-3}$) to a first approximation by the relation [13]

$$J_c = \frac{30[M^+ - M^-]}{D} \qquad \text{--- (1)}$$

where $M^+$ and $M^-$ are the magnetization values observed while sweeping the magnetic field up and down, respectively. Using $D = 2.3 \times 10^{-4}$ cm and ρ = x-ray density measured for each sample (Table – 1), values of $J_c$ (H) at different temperatures have been calculated using Eq. 1 for $La_{1.5}Nd_{0.5}Ca_1Ba_2(Cu_{1-x}Mo_x)_5O_z$ samples for x = 0.00, 0.01, 0.03 and 0.05 and the results are shown in Figures 6 (a & b).

## 4. DISCUSSION

The observed suppression of $T_c$ by Mo substitution in LNCBCuMO (Fig. 4) is explained on the basis of oxygen effect, assuming it is real. The oxygen content z, of a superconductor is directly related to the hole concentration p, which controls the superconductivity. It is evident from Fig. 4 (Table – 1) that $T_c$ decreases from 78 to 36 K as z decreases from 11.583 to 11.119 with a corresponding increase of x from 0.00 to 0.10. The trend is clear: the hole concentration p decreases as the samples are doped with a higher concentration of $Mo^{4+}$ for $Cu^{2+}$. The additional electrons contributed by $Mo^{4+}$ ions are expected to fill mobile holes in the $CuO_2$ planes, reducing conduction and eliminating superconductivity. As a matter of fact, the hole concentration in the sheets or planes ($p_{sh}$, Table – 1) decreases, from 0.225 to 0.085 with a corresponding decrease in $T_c$ from 78 to 36 K for x = 0.00 to 0.10. This suggests that Mo replaces Cu(2) plane sites.

The low Mo substitution of Mo at the Cu site in LNCBCuMO introduces point defects into the structure, which trap magnetic flux when subjected to external magnetic field. The increased Mo substitution at Cu-site beyond certain limit introduces lattice defects, which instead of trapping flux causes the scattering of charge carriers when subjected to external magnetic field. The flux pinning depends on flux trapping which is directly related to the intragrain $J_c$.

Figures 6 (a & b) show that $J_c$ decreases with increasing applied magnetic field (H) for $La_{1.5}Nd_{0.5}Ca_1Ba_2(Cu_{1-x}Mo_x)_5O_z$ samples with x = 0.00, 0.01, 0.03 and 0.05. The $J_c$ also decreases with increasing temperature. Figure 7 displays values of $J_c$ as a function of increasing Mo concentration at a particular field of 6 kOe and T = 5 K. The intragrain $J_c$ increases with increasing x up to x = 0.01, exhibiting a maximum value for $J_c$, and thereafter decreases with further increase in x for $0.01 \leq x \leq 0.05$. It is evident from Figure 7 that the $J_c$ value for x = 0.01 is nearly 2.5 times larger than the $J_c$ value of the pristine $La_{1.5}Nd_{0.5}Ca_1Ba_2Cu_5O_z$ (x = 0.00) sample at T = 5 K, while that of x = 0.03 is nearly the same as that of x = 0.00 at 5 K. As the increase in low Mo concentration from x = 0.00 to 0.01 produces increasing point defects up to x = 0.01, which increases intragrain $J_c$ as well as flux pinning up to x = 0.01 for a particular applied field (Fig. 7). Further increase in Mo-content beyond $x = 0.01$ up to $x = 0.05$, Mo plays the role of scattering charge carriers and producing lattice defects which increase with Mo-content leading to the decrease in flux pinning and $J_c$ up to $x = 0.05$ for a particular applied field (Fig. 7). This indicates that Mo doping for Cu at low concentration up to $x = 0.01$ produces point defects and increases both flux pinning and $J_c$ (Figs. 7 & 8), whereas for further increasing substitution of Mo ($0.01 < x < 0.05$), causes a carriers scattering by lattice defects leading to decrease in flux pinning and $J_c$ (Figs. 7 & 8). This indicates that Mo doping for Cu does enhance flux pinning in $La_{1.5}Nd_{0.5}Ca_1Ba_2(Cu_{1-x}Mo_x)_5O_z$ and that dopant acts as pinning centers. Similar $J_c$ behavior has been observed in Mo doped $YBa_2(Cu_{1-x}Mo_x)_3O_z$ [8].

The critical current density $J_c$ in the mixed state is related to the flux pinning force ($F_p$) by

$$F_p = J_c \times B \qquad \text{--- (2)}$$

where $F_p$ is bulk pinning force due to structural defects and B is the magnetic induction. Since we have determined $J_c$, it would be interesting to calculate $F_p$ for these systems assuming $J_c$ is perpendicular to B i.e. $F_p = J_cB$. Figure 8 (a) shows $F_p$ vs. H plots for $La_{1.5}Nd_{0.5}Ca_1Ba_2(Cu_{1-x}Mo_x)_5O_z$ for x = 0.00, 0.01, 0.03 and 0.05 at 5 K. For the sake of comparison and completeness we have also shown in Figure 8 (b), $F_p$ vs. H plots for Mo-doped Y-123 for x = 0.01 and 0.03 at 15 K. It is interesting to note that there is a similarity in plots for Mo-doped Y-123 and LNCBCuMO systems at x = 0.01 and 0.03.

A close look at Figure 8 (a) shows the increase in pinning force values for $La_{1.5}Nd_{0.5}Ca_1Ba_2(Cu_{1-x}Mo_x)_5O_z$ samples with increasing applied field, however, with increasing Mo-concentration the flux pinning decreases. For x = 0.01 sample, the maximum $F_p$ is obtained at comparatively low field. Hence, we can conclude at this point that the maximum current density (Figure 7) and flux pinning are obtained for lower concentration of Mo in $La_{1.5}Nd_{0.5}Ca_1Ba_2(Cu_{1-x}Mo_x)_5O_z$ type mixed oxide superconductors.

The flux creep model [14] accounts for the behavior of $J_c$ in magnetic fields. The field dependence of $J_c$ is given by [15]:

$J_c \propto H^{-n}$ (n = 0.5) --- (3)

In order to verify whether the presently studied samples satisfy the above relation, we plot in Figures 9 (a & b) ln $J_c$ vs. ln H obtained from Figures 6 (a & b). The slopes (-n) of the plots at T = 5 K are displayed in Figure 9 (c) as a function of Mo – concentration (x). It is evident from Figure 9 (c) that the values of n are within the range of expected value of 0.5 (n = ½). Thus the magnetic intragrain $J_c$ of Mo-doped LNCBCuMO samples shows a field dependence close to $J_c^m \propto H^{-0.5}$, which is interpreted as being indicative of bulk flux creep (Eq. 3).

## 5. CONCLUSIONS

Present studies on the $La_{1.5}Nd_{0.5}Ca_1Ba_2(Cu_{1-x}Mo_x)_5O_z$ clearly establish that the suppression of $T_c$ with increasing Mo-content mainly arises due to hole filling by Mo. Low Mo concentration does produce point defects. But further increase in Mo concentration, has adverse effects on overall properties of this system. At higher concentrations, Mo may be playing the role of scattering charge carriers and producing lattice defects, which reduces $T_c$ as well as $J_c$. Through comparison of the $J_c$ values and flux pining between pure $La_{1.5}Nd_{0.5}Ca_1Ba_2(Cu_{1-x}Mo_x)_5O_z$ (x = 0.00), and Mo-doped $La_{1.5}Nd_{0.5}Ca_1Ba_2(Cu_{1-x}Mo_x)_5O_z$ materials, it can be concluded that Mo-doping at low concentrations produces point defects which act as pinning center and promote enhancement of $J_c$.

**ACKNOWLEDGEMENT**

The present work has been carried out under the IUC-DAEF, Mumbai (INDIA) collaborative project (No. CRS-M-88) of Dr. D. G. Kuberkar. SR is thankful to IUC-DAEF for financial support.

**REFERENCES**

[1] D. G. Kuberkar, N. A. Shah, M. V. Subbarao, A. G. Joshi and R. G. Kulkarni Physica **B 259 – 261** (1999) 538

[2] D. G. Kuberkar, R. S. Thampi, N. A. Shah, Krushna Mavani, S. Rayaprol, S. K. Malik and W. B. Yelon, J. Appl. Phys. **89** (2001) 7657

[3] M. V. Subbarao, D. G. Kuberkar, G. J. Baldha and R. G. Kulkarni Physica **C 288** (1997) 57

[4] R. G. Kulkarni, D. G. Kuberkar, G. J. Baldha, G. K. Bichile Physica **C 217** (1993) 175

[5] A. G. Joshi, M. V. Subbarao, N. A. Shah, D. G. Kuberkar and R. G. Kulkarni Appl. Supercond. **6** (1998) 471

[6] P. H. Hor, R. L. Meng, Y. Q. Wang, L. Gao, Z. J. Huang, J. Bechtold, K. Forster, C. W. Chu, Phys. Rev. Lett. **58** (1987) 1891

[7] S. Rayaprol, Krushna Mavani, D. S. Rana, C. M. Thaker, R. S. Thampi, D. G. Kuberkar, R. G. Kulkarni and S. K. Malik, J. of Supercond. **15** (2002) 211 & S. Rayaprol, Krushna Mavani, C. M. Thaker, D. S. Rana, Keka Chakravorty, S. K. Paranjape, M. Ramanadham, Nilesh Kulkarni and D. G. Kuberkar. Pramana – Ind. J of Physics, **58** (2002) 877

[8] D. G. Kuberkar, J. A. Bhalodia, G. J. Baldha, I. A. Sheikh, Utpal Joshi and R. G. Kulkarni, Appl. Supercond. **3** (1995) 357

[9] R. G. Kulkarni, D. G. Kuberkar, A. G. Joshi and G. J. Baldha Physica C, **291** (1997) 25

[10] A. M. Kini, U. Geiser, H.C. I. Kao, K. D. Carson, H. H. Wang, M. R. Monaghan, J. M. Williams, Inorg. Chem **26** (1987) 1836

[11] Y. Tokura, J. B. Torrance, T. C. Huang and A. I. Nazzel Phys. Rev. **B 38** (1988) 7156

[12] C. P. Bean, Rev. Mod. Phys. **36** (1964) 31

[13] J. O. Willis, J. R. Cost, R. D. Brown, J. D. Thompson et. al IEEE Trans. Magn MAG-**25** (1989) 12

[14] M. Tinkham, Helv. Phys. Acta **61** (1988) 443

[15] J. Mannhart, P. Chaudhari, D. Dimos, C. C. Tsuei and T. R. Maguire Phys. Rev. Lett **61** (1988) 2476

**TABLE 1** Oxygen stoichiometry (z), oxygen stoichiometry per unit cell z' (=3/5*z), hole concentration (p), hole concentration in sheets ($p_{sh}$) and calculated density (ρ)

| Sample (x) | Z | z' | p | $p_{sh}$ | ρ (g / cm$^3$) |
|---|---|---|---|---|---|
| 0.00 | 11.583 (8) | 6.950 (3) | 0.300 | 0.225 | 10.4119 |
| 0.01 | 11.345 (8) | 6.807 (3) | 0.205 | 0.153 | 10.4856 |
| 0.03 | 11.322 (8) | 6.793 (3) | 0.195 | 0.146 | 10.4798 |
| 0.05 | 11.143 (8) | 6.686 (3) | 0.124 | 0.093 | 10.6056 |
| 0.10 | 11.119 (8) | 6.671 (3) | 0.114 | 0.085 | 11.0157 |

**TABLE 2** $T_c$ values (in K) obtained from measurements of resistance, magnetic susceptibility, and $J_c$ values at 5 K for 6 kOe

| Sample | Resistivity | | | Magnetic | $J_c$ |
|---|---|---|---|---|---|
| (x) | $T_c^{ON}$ | $T_c^{mid}$ | $T_c^{R=0}$ | $T_c^{ON}$ ($\chi_{mag}$) | (A / cm$^2$) |
| 0.00 | 80 (1) | 79 (1) | 78 (1) | 77 (1) | 1.2 X $10^6$ |
| 0.01 | 80 (1) | 76 (1) | 74 (1) | 75 (1) | 2.25 X $10^6$ |
| 0.03 | 78 (1) | 70 (1) | 66 (1) | 68 (1) | 1.5 X $10^6$ |
| 0.05 | 75 (1) | 70 (1) | 64 (1) | 65 (1) | 0.25 X $10^6$ |
| 0.10 | 60 (1) | 50 (1) | 39 (1) | 59 (1) | -- |

**FIGURE CAPTIONS**

Figure 1 Variation of lattice parameters (a & c) and volume of $La_{1.5}Nd_{0.5}Ca_1Ba_2(Cu_{1-x}Mo_x)_5O_z$ with x = 0.0 – 0.10

Figure 2 Oxygen content z and concentration of holes in planes ($p_{sh}$) with x for $La_{1.5}Nd_{0.5}Ca_1Ba_2(Cu_{1-x}Mo_x)_5O_z$

Figure 3 (a) Resistance vs. Temperature for $La_{1.5}Nd_{0.5}Ca_1Ba_2(Cu_{1-x}Mo_x)_5O_z$ at x = 0.00 – 0.10

(b) Magnetic susceptibility vs. temperature for $La_{1.5}Nd_{0.5}Ca_1Ba_2(Cu_{1-x}Mo_x)_5O_z$ for x = 0.00 – 0.10

Figure 4 $T_c$ vs. x for $La_{1.5}Nd_{0.5}Ca_1Ba_2(Cu_{1-x}Mo_x)_5O_z$

Figure 5 Typical magnetic hysteresis loops for $La_{1.5}Nd_{0.5}Ca_1Ba_2(Cu_{1-x}Mo_x)_5O_z$

Figure 6 Magnetic field dependence of $J_c$ for $La_{1.5}Nd_{0.5}Ca_1Ba_2(Cu_{1-x}Mo_x)_5O_z$ (a) at various concentrations (x) for T = 5 K and (b) at T = 40 K for x = 0.01 and 0.03

Figure 7 $J_c$ vs. x for $La_{1.5}Nd_{0.5}Ca_1Ba_2(Cu_{1-x}Mo_x)_5O_z$

Figure 8 Variation of pinning force ($F_p$) with applied field H (a) for $La_{1.5}Nd_{0.5}Ca_1Ba_2(Cu_{1-x}Mo_x)_5O_z$ at 5 K for x = 0.00 – 0.05 (b) for $YBa_2(Cu_{1-x}Mo_x)_3O_z$ at 15 K for x = 0.01 and 0.03

Figure 9 (a) Plot of ln $J_c$ vs. ln H for x = 0.00 – 0.05 at 5 K

(b) Plot of ln $J_c$ vs. ln H for x = 0.01 and 0.03 at 40 K

(c) Plot of n (slope of ln $J_c$ vs. ln H plot) vs. x at 5 K

Figure 1

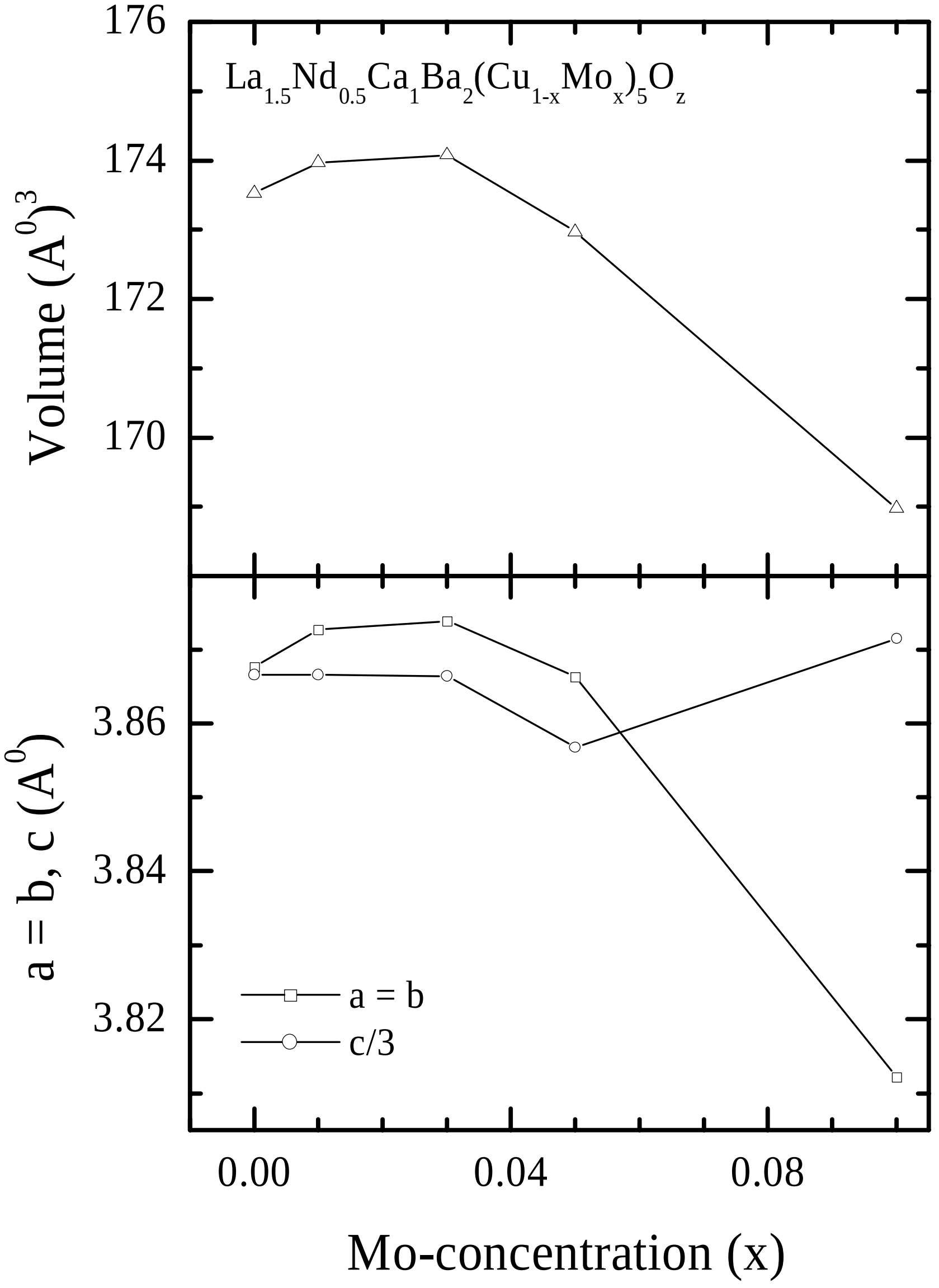

Figure 2

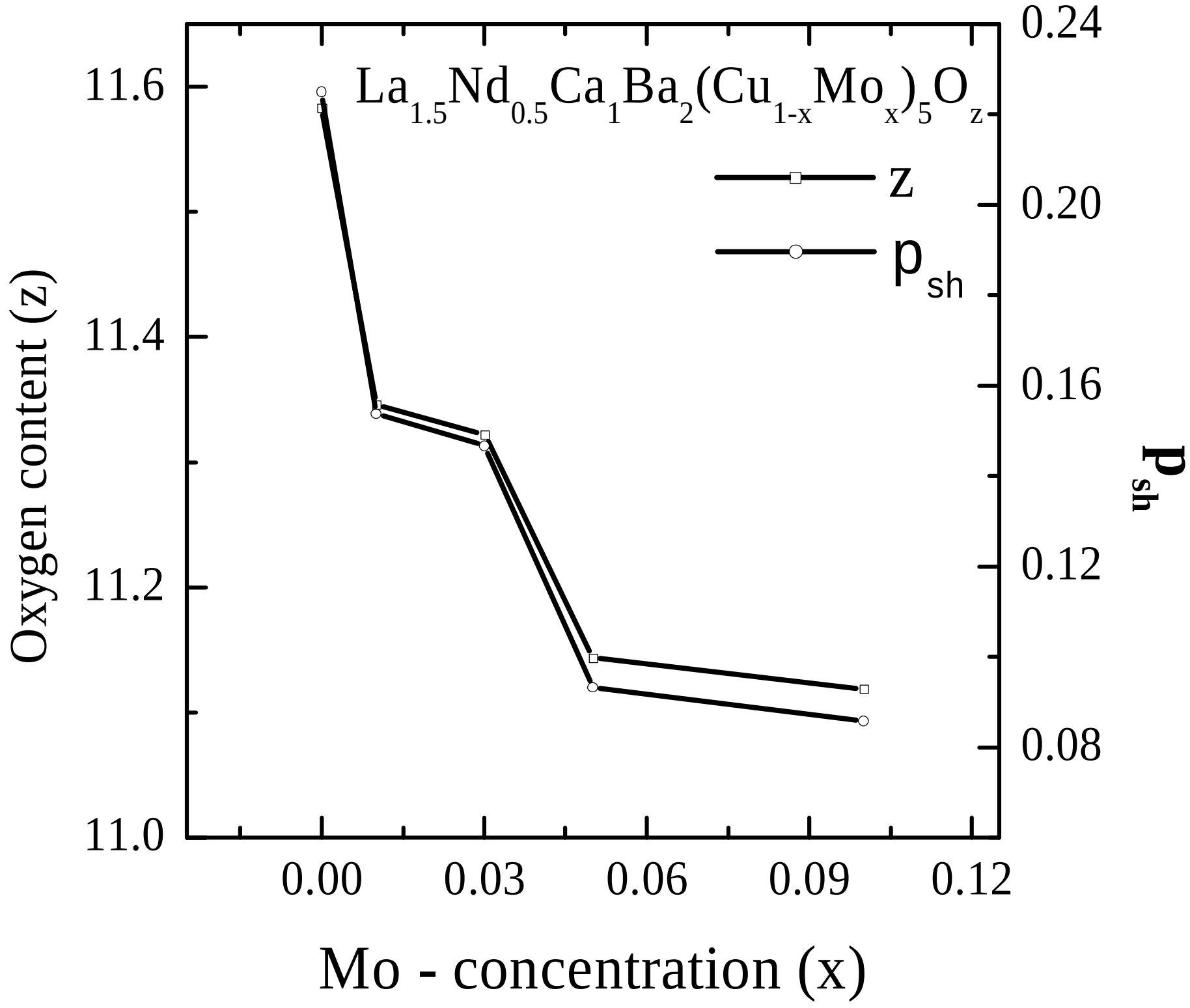

Figure 3 (a)

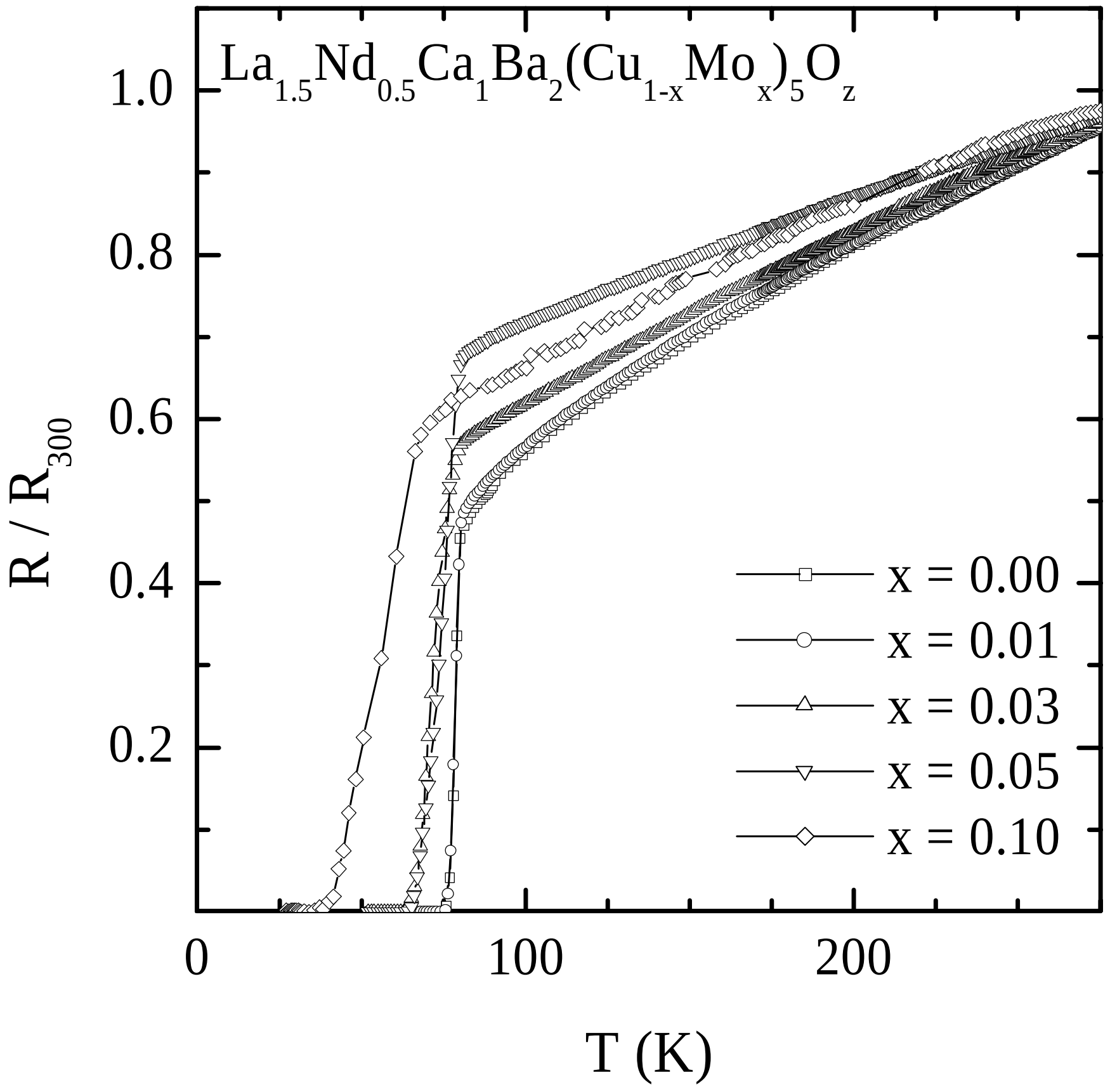

Figure 3 (b)

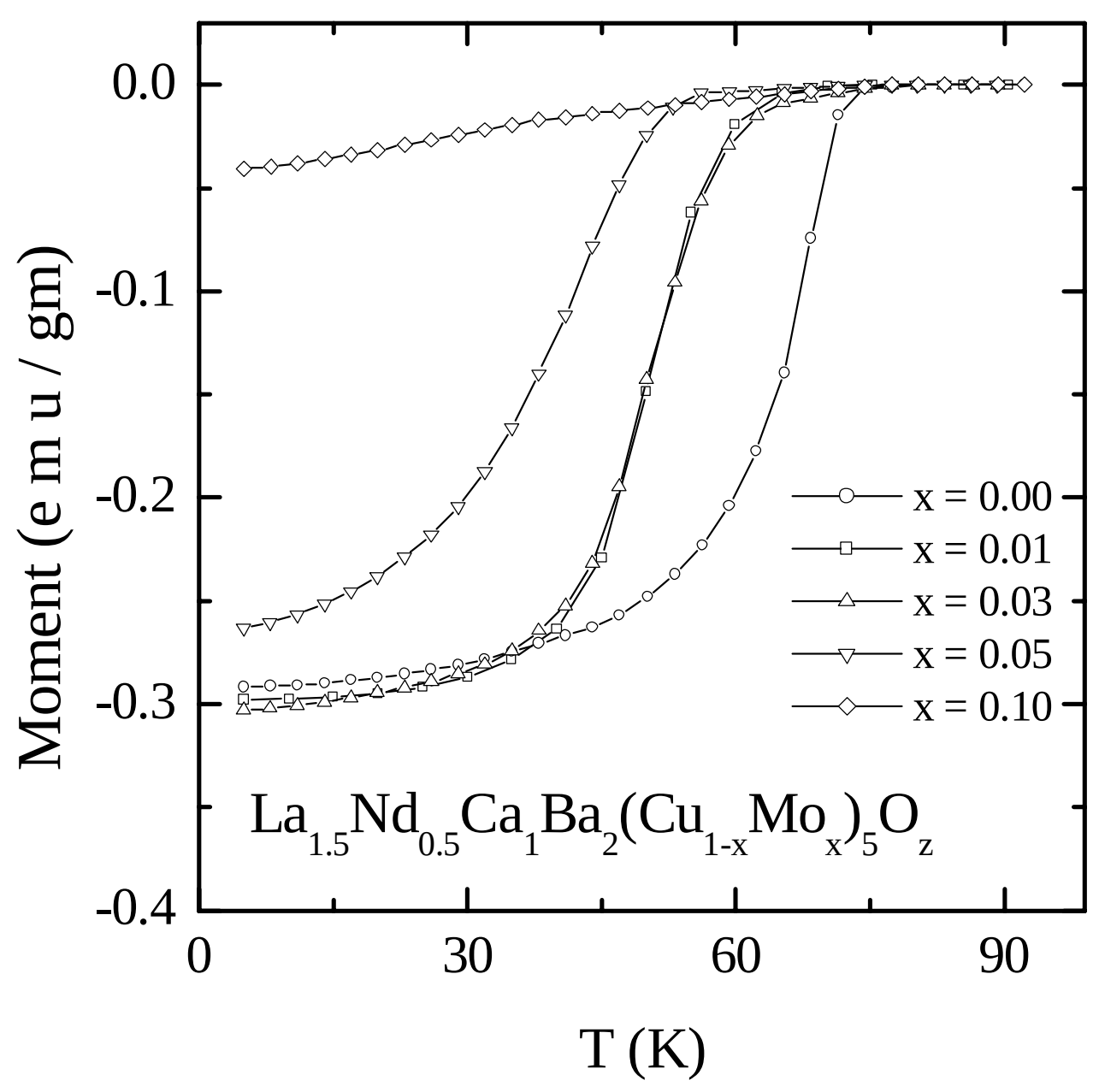

Figure 4

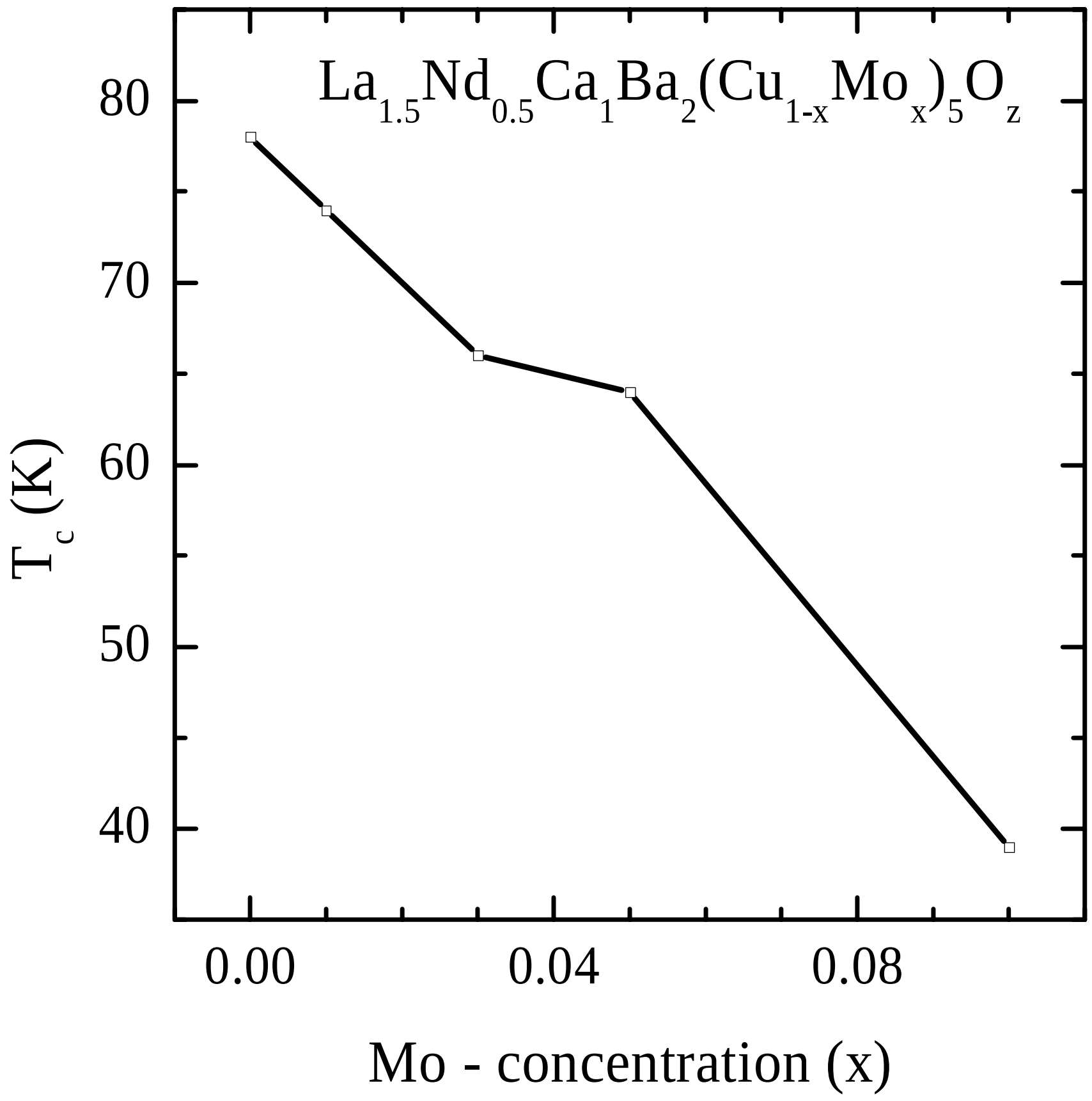

Figure 5

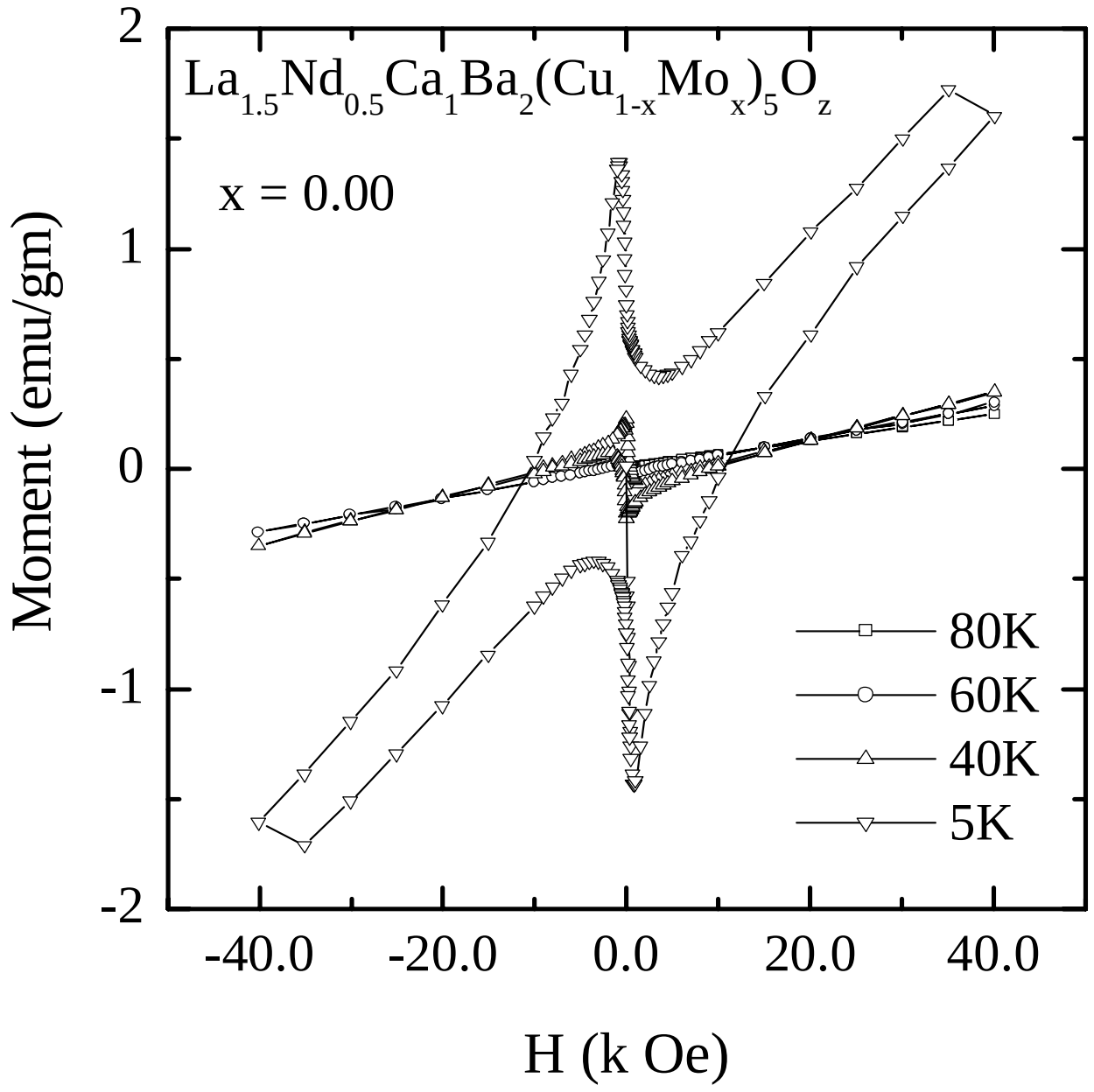

Figure 6 (a)

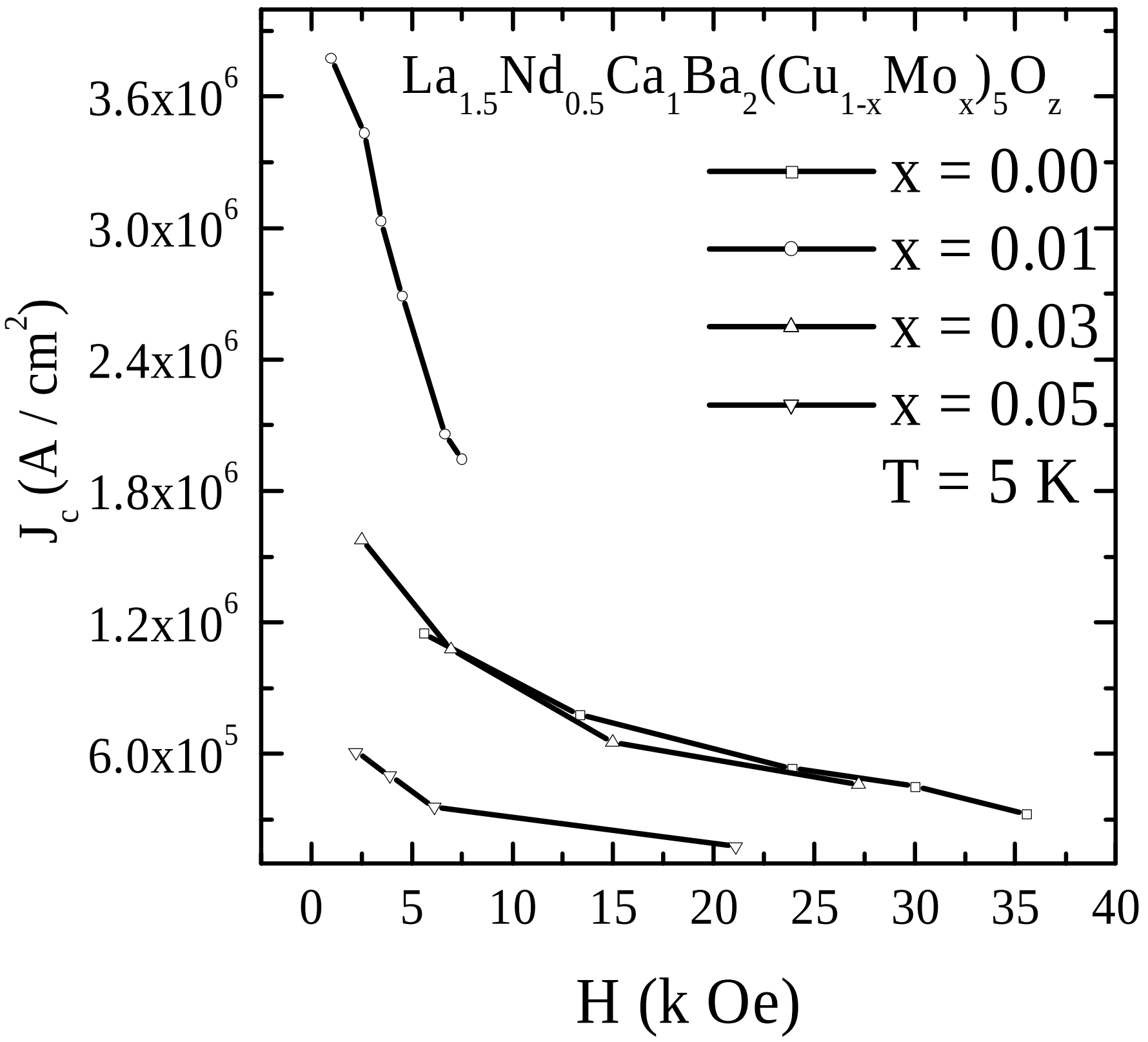

Figure 6 (b)

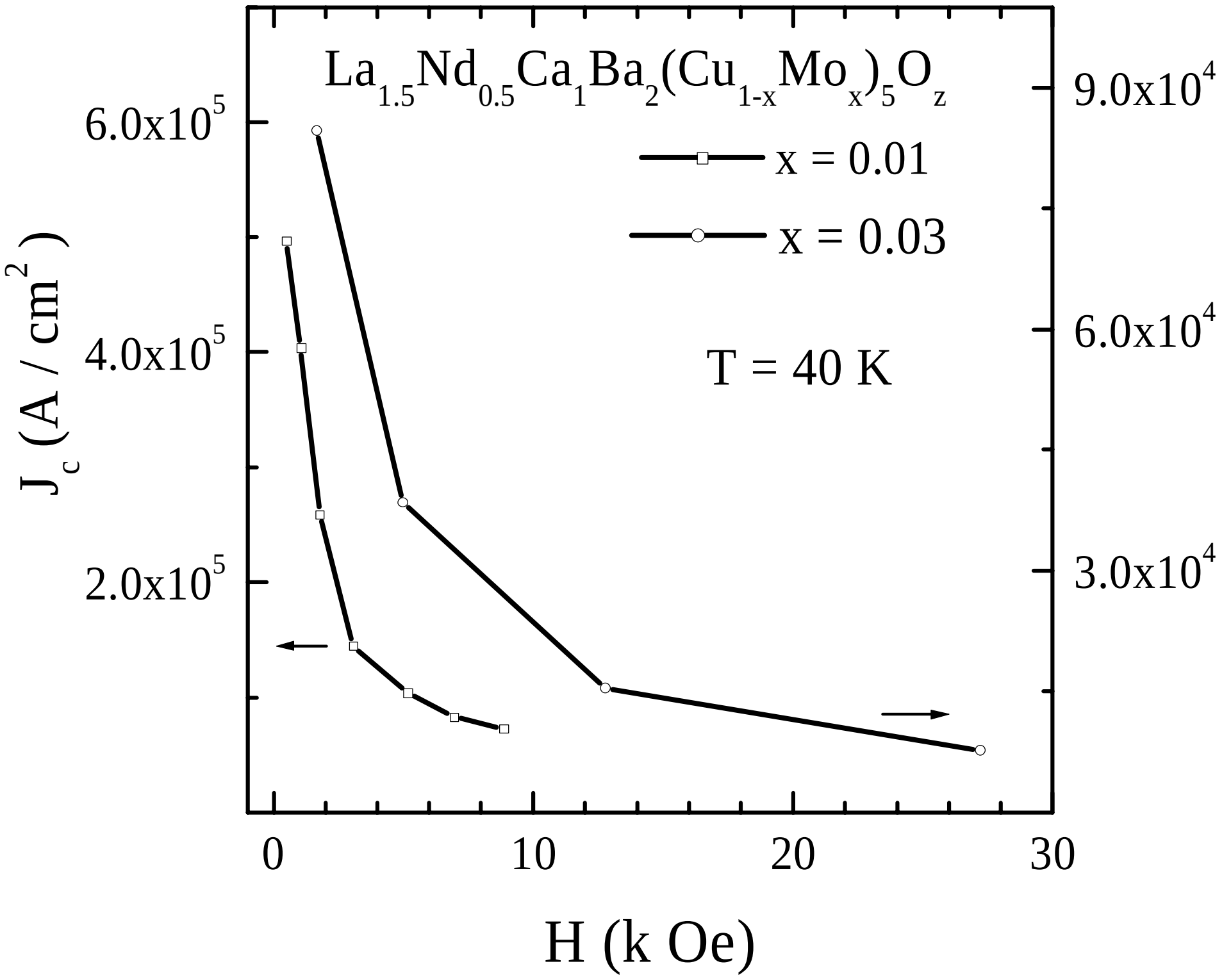

Figure 7

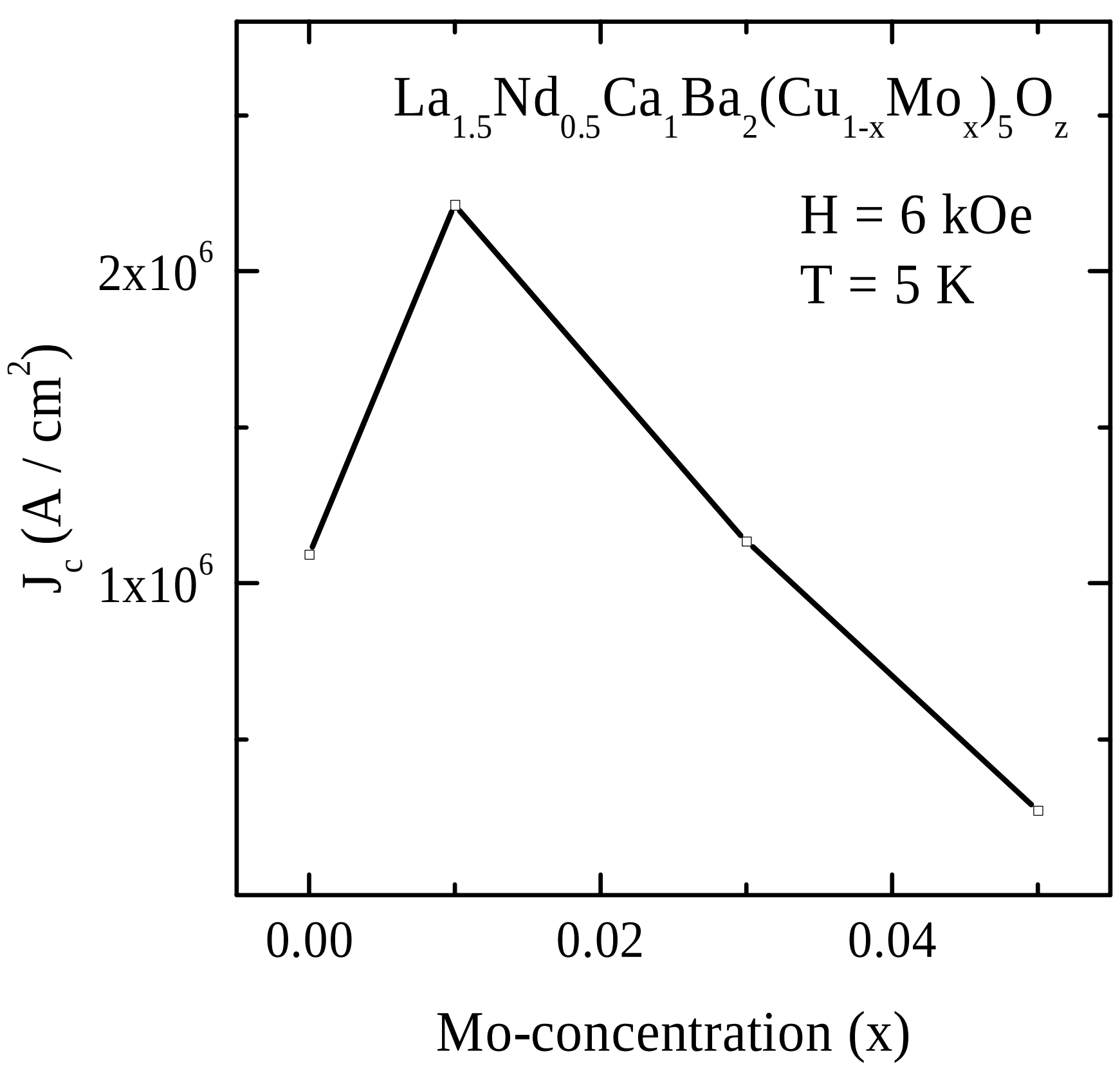

La1.5Nd0.5Ca1Ba2(Cu1-xMox)5Oz
H = 6 kOe
T = 5 K
2x10^6
1x10^6
Jc (A / cm^2)
0.00
0.02
0.04
Mo-concentration (x)

Figure 8 (a)

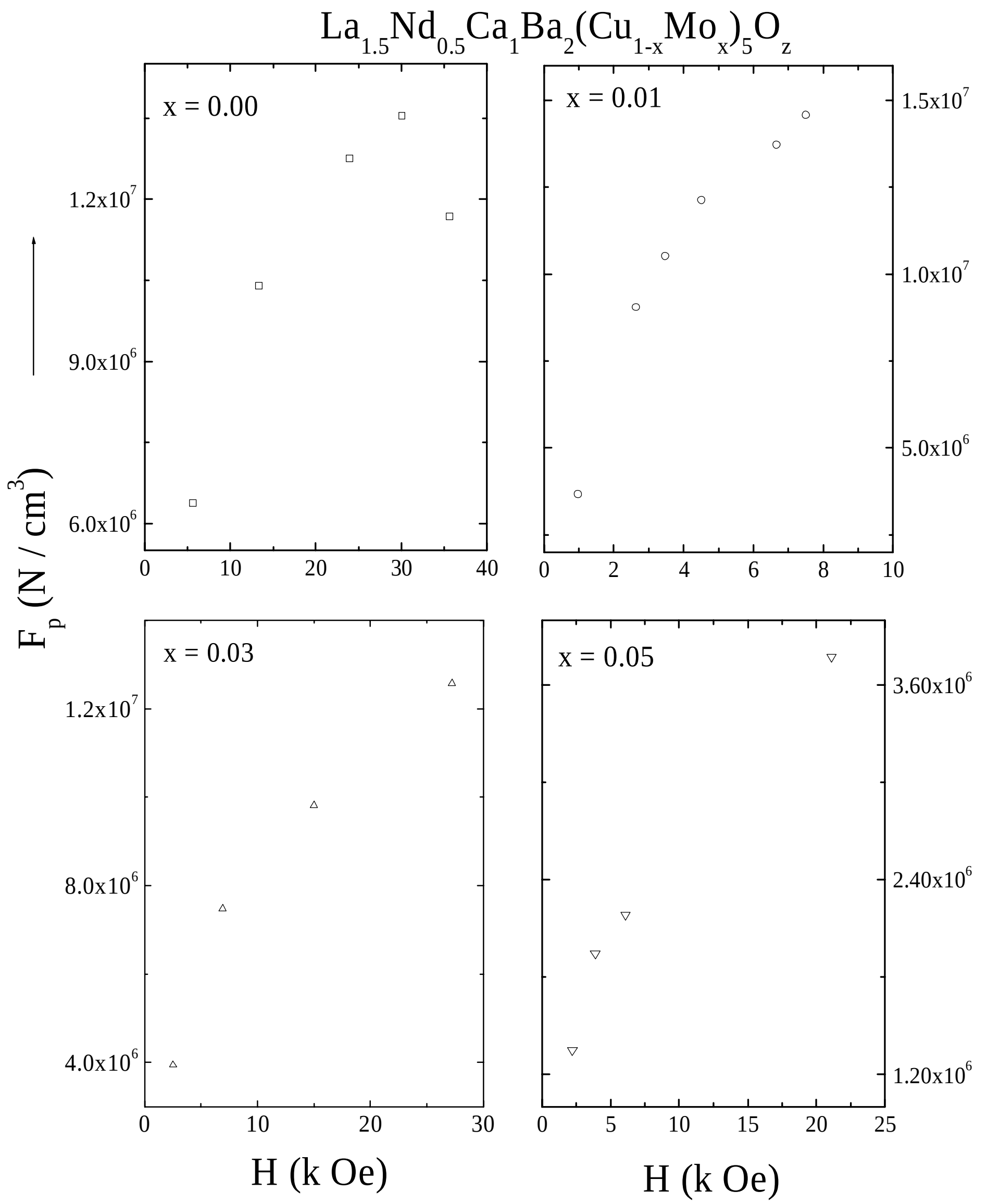

$La_{1.5}Nd_{0.5}Ca_1Ba_2(Cu_{1-x}Mo_x)_5O_z$
x = 0.00
x = 0.01
x = 0.03
x = 0.05
1.2x10^7
9.0x10^6
6.0x10^6
1.5x10^7
1.0x10^7
5.0x10^6
1.2x10^7
8.0x10^6
4.0x10^6
3.60x10^6
2.40x10^6
1.20x10^6
0 10 20 30 40
0 2 4 6 8 10
0 10 20 30
0 5 10 15 20 25
$F_p$ (N / cm^3)
H (k Oe)
H (k Oe)

Figure 8 (b)

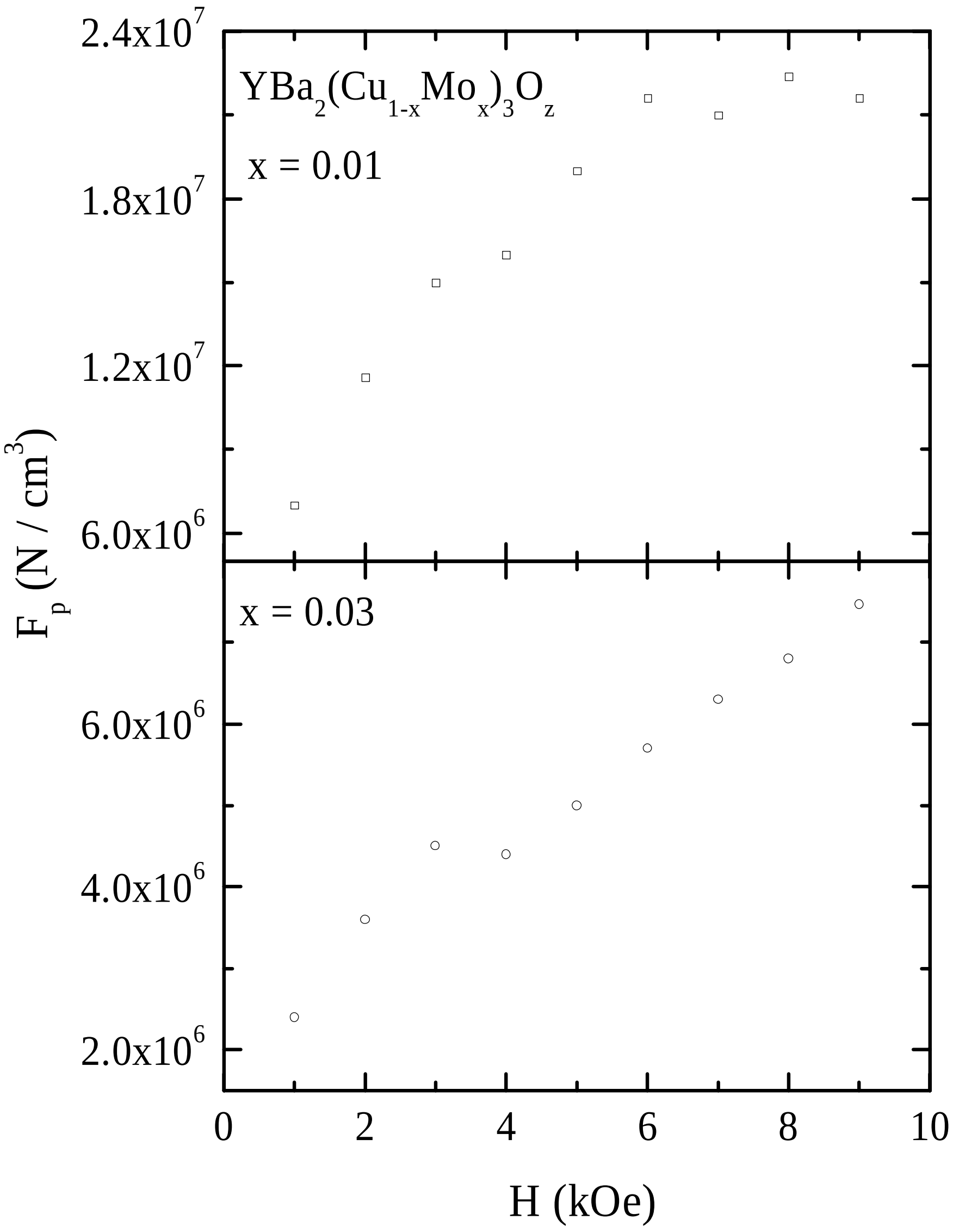

Figure 9 (a)

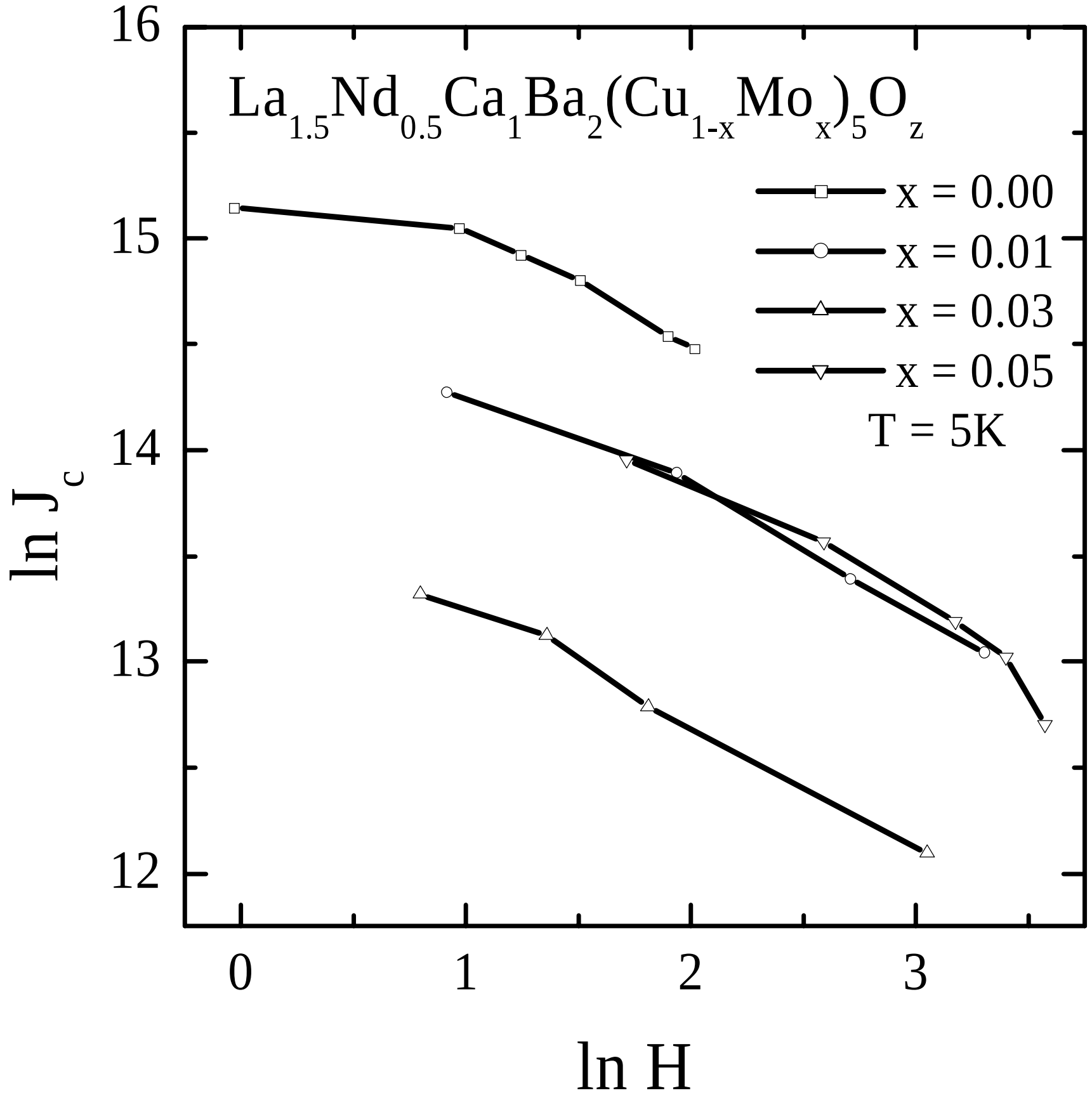

Figure 9 (b)

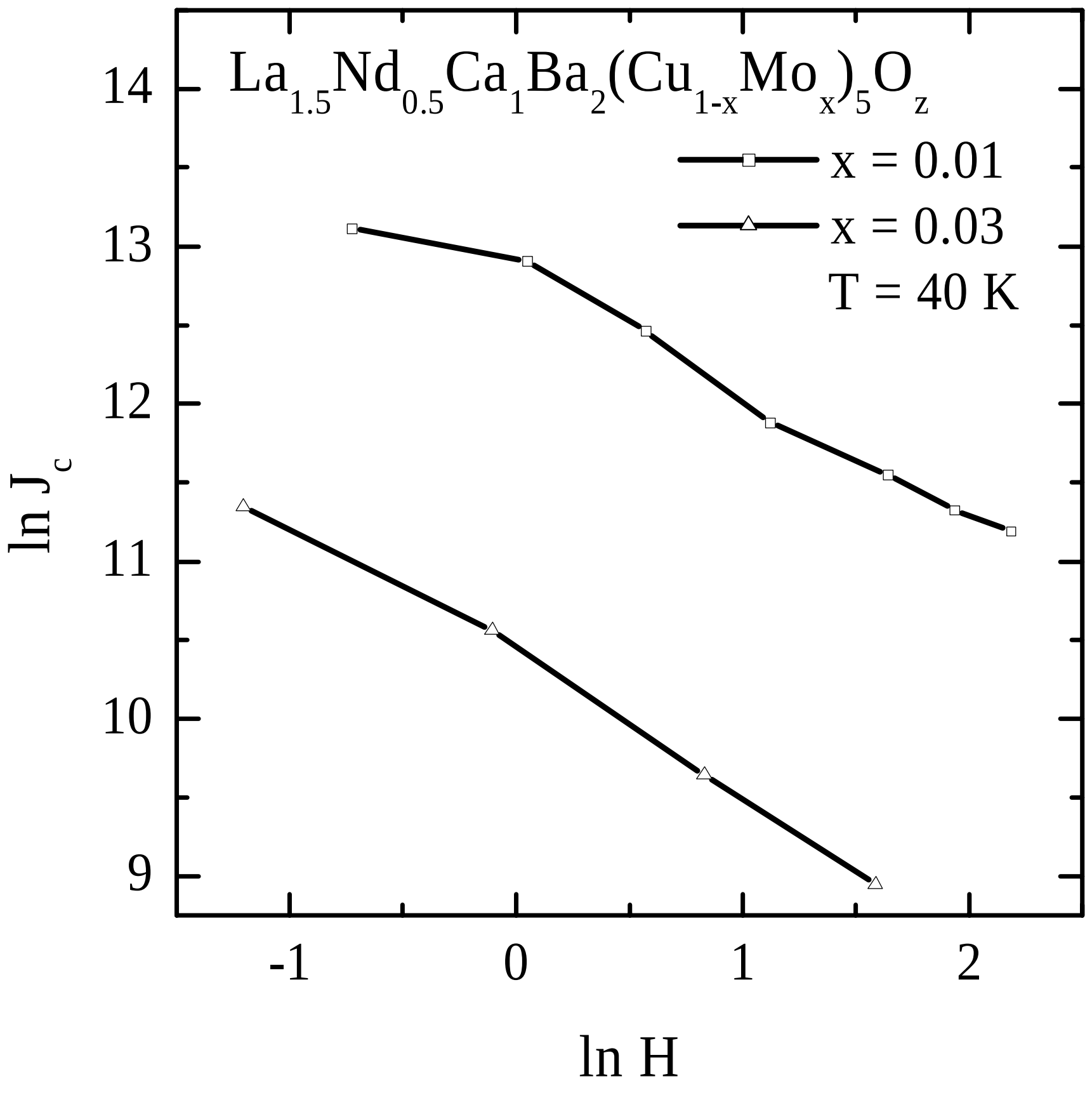

Figure 9 (c)

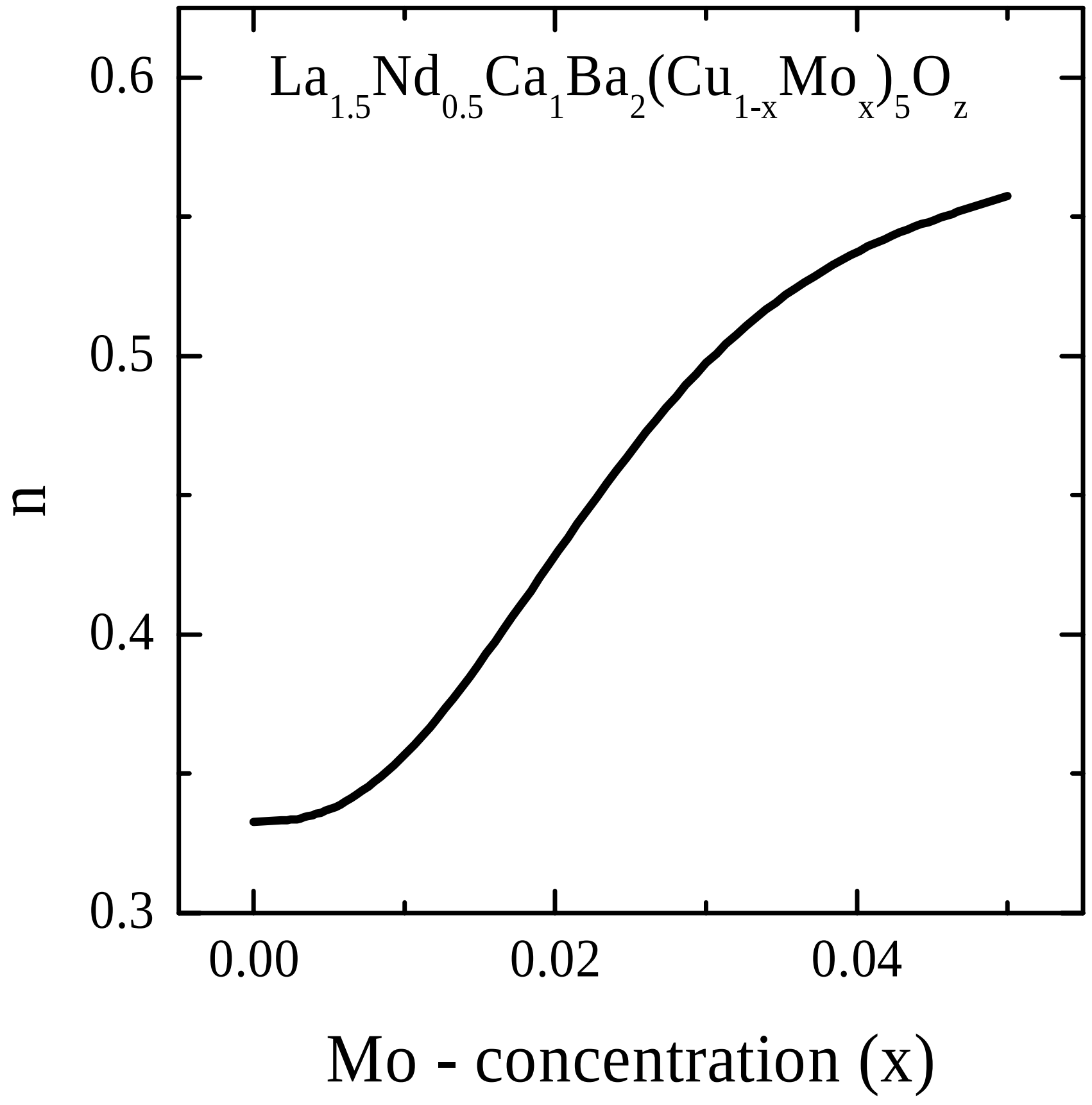